\documentclass[twocolumn,prb,aps,epsf,amsmath,amssymb,showpacs]{revtex4}
\usepackage{graphicx}
\usepackage{dcolumn}
\usepackage{bm}

\usepackage{comment}
\usepackage[dvipsnames]{xcolor}
\usepackage{soul}

\begin{document}
\baselineskip=13pt

\preprint{working draft}

\title{ Dynamics of Supersolid state: normal fluid, superfluid, and supersolid velocities}
\author{ Wayne M. Saslow} 
\email{wsaslow@tamu.edu}
\affiliation{ Department of Physics, Texas A\&M University, College Station, 
TX 77843-4242}
\date{\today}

\begin{abstract}
Landau's excitation-based argument for superfluids -- that at temperature $T=0$ the normal fluid density $\rho_{n}$ is zero -- should also apply to supersolids.  Further, for a total mass density $\rho$, Leggett argues that the superfluid fraction $\rho_{s}/\rho<1$.  These arguments imply that there is a missing mass. We attribute this to a supersolid density $\rho_{L}$, with $\rho_{L}\equiv \rho-\rho_{s}-\rho_{n}$, and a momentum-bearing supersolid velocity $v_{Li}$.  Using Onsager's irreversible thermodynamics we derive the macroscopic dynamical equations for this system.  We find that  $v_{Li}$ is subject to the force of elasticity, to the negative gradient of the chemical potential per mass $\mu$ (as for the superfluid velocity $v_{si}$), and to drag against the normal fluid (leading to the interpretation of $L$ as lattice).  Thus both the superfluid and supersolid components are associated with the ground state.  The normal modes for such a system have a crossover in frequency, above which the normal fluid velocity $v_{ni}$ is an independent variable and below which it is locked to $v_{Li}$.  For an isotropic lattice we study both the transverse response and longitudinal response.  The ring geometry for atomic gas supersolid states may provide a geometry for testing these predictions. 
\end{abstract}


\maketitle

\section{Introduction}
\label{s:Intro}
To explain the unusual flow properties of liquid $^{4}$He below the $\lambda$ temperature of 2.17 K, Landau (1941) developed the theory of the superfluid state, having two components with mass -- superfluid and normal fluid -- and corresponding momentum-bearing velocities.\cite{Landau41}  
In 1969 Andreev and Lifshitz (AL)\cite{AL69} considered the possibility of superfluid flow in solids, attributing inertial mass to the Galilean superfluid velocity $v_{si}$ and the Galilean normal fluid $v_{ni}$ velocity, but not to $\partial_{t}u_{i}$, where $u_{i}$ is the lattice displacement.\cite{OrdSolid}  
In 1970 Chester, in the context of the relation between Bose-Einstein condensation and superfluidity, suggested that one could have a solid phase based on vacancies, presumably with both superfluid and normal fluid components.\cite{Chester70}  

In his theory of a hypothetical solid displaying superflow, Leggett (1970) considered at $T=0$ what we call ``phase-flow'', a dissipation-free flow like what occurs for electrons in atomic states with non-zero angular momentum.\cite{Leggett70,AmperianCurrent,macrocurrents}  He observed that, for an annulus of radius $R$ rotating at angular velocity $\omega$, and thus with velocity $v=\omega R$, in the rotating frame (where the Hamiltonian is fixed) on going around the annulus there is a net phase change $\Delta \Phi=mv(2\pi R)/\hbar$.  

By a variational method Leggett showed that to adjust to this net phase change the atomically-varying phase of the many-body wavefunction lowers the kinetic energy by preferentially developing in regions of lower local atomic density.  He interpreted the decreased kinetic energy as a Nonclassical Rotational Inertia (NCRI) that even at $T=0$ is less than the classical value: equivalently, a superfluid fraction $\rho_{s}/\rho<1$.  In this theory $v_{si}$ is defined relative to the lattice, and is not a Galilean velocity.  With solid $^{4}$He in mind, explicit calculations for $\rho_{s}$ were based on this idea.\cite{FernandezPuma74,GRS07,Saslow12}  
In describing boson condensation in a cell model (1971), Mullin coined the term ``supersolid''.\cite{Mullin71}  

Despite early promise of supersolid behavior in solid $^{4}$He in NCRI studies,\cite{KimChan04} these results were later reinterpreted as shear stiffening.\cite{KimChan12}  For a review see Ref.~\onlinecite{ChanReview14}.  On the other hand, developments in confined atomic gases, which are tunable by atomic variety and combinations, and by density, temperature, and magnetic field, have yielded evidence of supersolid order.  Initial studies had a confining potential that gives the normal modes a gap for $k=0$.\cite{Leonard17,Li17,Tanzi19A,Bottcher19,Chomaz19}  Later studies have involved self-confining interactions between the atoms, which is of more direct relevance to the present work.\cite{Tanzi19B,Guo19,Lev21}  Quite recently both atomic lattices and quantum vortices, produced by ``magnetic stirring'', have been observed, providing evidence both of solid order and of superfluid order.\cite{Ferlaino24}  

Landau's reasoning about the superfluid state -- that the normal fluid density $\rho_{n}$ is due to thermal excitations -- also holds for the supersolid state, so at $T=0$ we have $\rho_{n}=0$.  Applying the reasoning of both Landau and Leggett, we conclude that at $T=0$ --- and generally at low enough temperatures --- there is a missing inertial mass.  We attribute this to the supersolid fraction, with supersolid velocity $v_{Li}$ and supersolid density $\rho_{L}\equiv \rho-\rho_{s}-\rho_{n}$.  This implies that the supersolid state has three momentum-bearing velocities, whereas the superfluid state has only two.  As with AL, the present work applies Onsager's irreversible thermodynamics to obtain the macroscopic equations of motion. This work should be relevant to atomic gases, particularly in an annular geometry. 

\section{Results} 
\label{s:Results}

Of the two possible transformation properties for $v_{si}$ -- the Galilean one of AL and the lattice-frame one of Leggett -- we find that only the Galilean $v_{si}$ has non-dissipative uniform superflow.  We therefore take $v_{si}$ to be Galilean. 

We find equations of motion for the average mass density $\rho$, $v_{si}$, and entropy density $S$ that agree with Landau for a superfluid.\cite{Landau41}. We also find an equation of motion for the current density $j_{i}$ that agrees with AL for a supersolid.\cite{AL69}  
The theory further predicts that the non-momentum-bearing velocity $v_{Ei}\equiv\dot{u}_{i}$ equals $v_{Li}$ up to diffusive terms.

The equation of motion for $v_{Li}$, on neglecting diffusive terms, has three contributions. \\
\indent{} a. The first term, not unexpectedly, is the acceleration due to elastic forces. \\  
\indent{} b. The second term is due to frictional drag (with drag time $\tau$) between the supersolid $v_{Li}$ and the normal fluid $v_{ni}$, as if the supersolid is a net through which the normal fluid moves.\cite{inertialdrag}  This term dominates at low frequencies, causing $v_{Li}\approx v_{ni}$.  Then one can think of an effective normal fluid density $\tilde{\rho}_{n}\equiv\rho_{n}+\rho_{L}$ moving with $v_{ni}$ with the resulting equations being very much like those of AL. \\  
\indent{} c. The third term is an acceleration given by the negative gradient of the chemical potential per unit mass, just as for the superfluid component.  This means that both the superfluid and the supersolid components are associated with the ground state. 

We have studied the normal modes of this system, both at low and high frequencies relative to $\tau^{-1}$.\\ 
\indent{} a. Transverse motion does not involve the superfluid.  At short times drag causes the supersolid and the normal fluid to move together.  Once they move together, there is an elasticity-driven mode.  At frequencies $\omega\tau\gg1$, there are two coupled modes of the supersolid and the normal fluid. \\
\indent{} b. Longitudinal motion involves all three velocities.  At low frequencies, because of drag the normal fluid follows the supersolid to give two coupled modes.  At high frequencies, where drag may be neglected, we expect three propagating modes, but for a perfect crystal the equations tell us otherwise.  One of the modes, which the equations of motion show is associated with changes in the chemical potential, is at zero frequency.  It corresponds to $v_{ni}=0=j_{i}$, but with non-zero superfluid $v_{si}$ and supersolid $v_{Li}$ together giving $j_{i}=0$.  For a slightly imperfect lattice and no viscous drag the velocity of this mode can be obtained by perturbation theory.\cite{WMSUnpublished}  The other two modes have velocities given by a quadratic equation but with no obvious physical interpretation.  




\section{Thermodynamics of a Supersolid}
\label{s:Thermo}

With $j_{si}=\rho_{s}(v_{si}-v_{ni})$ and $j_{Li}=\rho_{L}(v_{Li}-v_{ni})$ in the normal fluid frame, where $\rho_{s}$ and $\rho_{L}$ typically are tensors, a boost to the general frame gives 
\begin{equation}
j_{i}=\rho v_{ni}+ j_{si}+ j_{Li}=\rho_{s}v_{si}+\rho_{n}v_{ni}+\rho_{L}v_{Li}.
\label{j2}
\end{equation}

We must modify AL by including the effect of $j_{Li}$ and $v_{Li}$.  Following AL we employ energy density $E$, entropy density $S$, chemical potential (per unit mass) $\mu$, mass density $\rho$, and pressure $P$.  The stress density is $\lambda_{ik}$ and the strain is $w_{ik}=\partial_{i}u_{k}$.  Consistent with the classical kinetic energy for three masses and three velocities, we take as thermodynamic relations 
\begin{eqnarray}
dE{\hskip-0.15cm}&=&{\hskip-0.15cm}TdS+\mu d\rho+\lambda_{ik}dw_{ik}+v_{ni}dj_{i}+j_{si}dv_{si}+j_{Li}dv_{Li},\qquad
\label{dE123}\\
E{\hskip-0.15cm}&={\hskip-0.15cm}&TS-P +\lambda_{ik}w_{ik} + \mu \rho+v_{ni}j_{i}+j_{si}v_{si}+j_{Li}v_{Li},
\label{E123}\\
0{\hskip-0.15cm}&=&{\hskip-0.15cm}SdT-dP+\rho d\mu +w_{ik}d\lambda_{ik}+j_{i}dv_{ni}+v_{si}dj_{si}\cr
&&+v_{Li}dj_{Li}.  
\label{G-Duh123} 
\end{eqnarray}
In each of the above equations the last term is new.  Eq.~\eqref{G-Duh123} is the Gibbs-Duhem relation, which we will use later. 

The differentials in $dE$ indicate we will need equations of motion for $S$, $\rho$, $j_{i}$, $w_{ik}$, $v_{si}$, and the new variable $v_{Li}$.  Moreover, the thermodynamic ``forces'' will involve their thermodynamic conjugates $T$, $\mu$, $\lambda_{ki}$, $v_{ni}$, $j_{si}$, and $j_{Li}$.  

We consider an isotropic lattice, for which, with the symmetrized form $u_{ij}=\tfrac{1}{2}(w_{ij}+w_{ki})$, the strain-dependent free energy density is $F=\tfrac{1}{2}\lambda u_{ii}^{2}+\bar{\mu}u_{ik}^{2}$, where $\lambda$ and $\bar{\mu}$ are the Lam\'e constants.\cite{LLElasticity}  This can be rewritten as $F=\bar{\mu}(u_{ik}-\tfrac{1}{3}u_{ll}\delta_{ik})^{2}+\tfrac{1}{2}K u_{ik}^{2}$, where $K=\lambda+\tfrac{2}{3}\bar{\mu}$ is the modulus of compression (or bulk modulus) and $\bar{\mu}$ is the modulus of rigidity (or shear modulus).  Then
\begin{equation}
\lambda_{ik}\equiv\frac{\partial F}{\partial u_{ik}}=K\delta_{ik}u_{ll}+2\bar\mu(u_{ik}-\tfrac{1}{3}\delta_{ik}u_{ll}).
\label{lambdaik1}
\end{equation} 

For transverse motion, $\partial_{i}u_{i}=0$, so 
\begin{equation}
\partial_{i}\lambda_{ik}=\bar{\mu}\nabla^{2}u_{i}= K_{\perp}\nabla^{2}u_{i}, \quad K_{\perp}\equiv \bar{\mu}.
\label{Kperp}
\end{equation}

For longitudinal motion, $\partial_{i}\partial_{k}u_{k}=\nabla^{2}u_{i}$, so 
\begin{equation}
\partial_{k}\lambda_{ik}=(K+\tfrac{4}{3}\bar{\mu})\nabla^{2}u_{i}= K_{\parallel}\nabla^{2}u_{i}, \quad K_{\parallel}\equiv K+\tfrac{4}{3}\bar{\mu}.
\label{Kpar}
\end{equation}

\section{Dynamics of a Supersolid}
\label{s:Dynamics}
In irreversible thermodynamics one writes the rate of heat production $R>0$ as the sum of a divergence and products of the known thermodynamic forces ${\cal F}_{\alpha}$ with corresponding sources or fluxes ${\cal J}_{\alpha}$,\cite{Onsager1} which often are gradients of the differential coefficients in \eqref{dE123}.  Near equilibrium the ${\cal J}_{\alpha}$ are proportional to the ${\cal F}_{\alpha}$ with a matrix whose form is determined by the symmetry of the thermodynamic state and by the condition that $R>0$.   

We take the equations of motion to be\cite{FN4}
\begin{eqnarray}
\dot{\rho}+\partial_{i} j_{i}&=&0,
\label{drhodt1}\\
\frac{\partial j_{i}}{\partial t}+\frac{\partial \Pi_{ik}}{\partial x_{k}}&=&0, 
\label{djdt1}\\
\dot{u}_{i}&\equiv &v_{Ei}=U_{i},
\label{udot1}\\
\dot{v}_{si}&=& -\partial_{i}\varphi, 
\label{dvsdt1}\\
\dot{v}_{Li}&=& Y_{i},
\label{dvLdt1}\\
\dot{S}+\partial_{i}\Big[Sv_{ni}+\frac{q_{i}}{T}\Big]&=&\frac{R}{T} (R>0), 
\label{dSdt1}\\
\dot{E}+\partial_{i}Q_{i}&=&0.
\label{Edot}
\end{eqnarray}

On applying irreversible thermodynamics (see Appendix) we find the sources and fluxes $j_{i}$, $\Pi_{ik}$, $\varphi$, $U_{i}$, $Y_{i}$, $q_{i}$, and $R$.  We have $j_{i}$ given by \eqref{j2} and, for a perfect lattice, $\Pi_{ik}=-\lambda_{ik}$, which we compute using  
$K_{\perp}=\bar{\mu}$, $K_{\parallel}=K+\tfrac{4}{3}\bar{\mu}$, and $\bar{K}=K+\tfrac{1}{3}K_{\perp}$, so that
\begin{equation}
\frac{\partial j_{i}}{\partial t}=\frac{\partial \lambda_{ik}}{\partial x_{k}}=\bar{K}\partial_{i}\partial_{k}u_{k}+K_{\perp}\nabla^{2}u_{i}. 
\label{stress}
\end{equation}  

Then, neglecting diffusion, we find (see Appendix) 
\begin{eqnarray}
\dot{u}_{i}&= &U_{i} \approx v_{Li},
\label{udot*}\\
\dot{v}_{si}& \approx &-\partial_{i}\mu, 
\label{dvsdt1L}\\
\dot{v}_{Li}= Y_{i}& \approx &-\partial_{i}\mu +\frac{\bar{K}}{\rho_{L}}\partial_{i}\partial_{k}u_{k}+\frac{K_{\perp}}{\rho_{L}}\nabla^{2}u_{i}\cr
&&\quad -\tau^{-1}(v_{Li}-v_{ni}), 
\label{dvLdt1L}\\
\dot{S}+\partial_{i}(Sv_{ni})& \approx &0. 
\label{dSdt1L}
\end{eqnarray}
In the above, for an isotropic solid $\tau^{-1}$ represents a diagonal matrix with one parallel and two degenerate perpendicular components. 

\section{Transverse Modes ($\perp$)}
\label{s:transverse}
We consider consider small deviations from equilibrium of the form $e^{(-i\omega t + ikz)}$.  For transverse motion, \eqref{drhodt1} implies that $\rho$ does not vary, and \eqref{dSdt1L} implies that $S$ does not vary.  Therefore no thermodynamic quantity varies, so \eqref{dvsdt1L} gives $v_{si}=0$.  Hence transverse motion involves only transverse $v_{Li}$ and transverse $v_{ni}$.  

Eqs.~\eqref{drhodt1} for $\partial_{t}{j}_{i}$ and \eqref{dvLdt1L} for $\dot{v}_{Li}$ give
\begin{eqnarray}
0&=&-i\omega(\rho_{n\perp}v_{ni}+\rho_{L\perp}v_{Li}) + K_{\perp} k^{2} u_{i}.
\label{djdtA}\\
-i\omega v_{Li} &=& -\rho_{L\perp}^{-1} K_{\perp} k^{2} u_{i} - \tau_{\perp}^{-1}(v_{Li}-v_{ni}).
\label{dvLdtA}
\end{eqnarray}
The two transverse directions have the same mode frequencies, so we drop the index $i$.   
Taking $d/dt\equiv-i\omega$ on each of the above equations and using $-i\omega u_{}=v_{E}\approx v_{L}$ gives
\begin{eqnarray}
\omega^{2}\rho_{n\perp}v_{n}&=&(-\omega^{2}\rho_{L\perp}+ K_{\perp} k^{2}) v_{L}, \qquad
\label{djdtB2}\\
(\omega^{2} - \rho_{L\perp}^{-1} K_{\perp} k^{2} +  i\omega\tau_{\perp}^{-1})v_{L} &=& i\omega\tau_{\perp}^{-1}v_{n}.
\label{dvLdtB2}
\end{eqnarray}
With 
\begin{equation}
r_{\perp} \equiv \frac{\rho_{L\perp}}{\rho_{n\perp}}, \quad c_{\perp} \equiv \sqrt{\frac{K_{\perp}}{\rho_{\perp}}}, 
\label{consts}
\end{equation}
cross-multiplication of \eqref{djdtB2} and \eqref{dvLdtB2} leads to 
\begin{equation}
k^{2}=\frac{\omega^{2}}{c^{2}_{\perp}} \frac{\omega+i\tau_{\perp}^{-1}(1+r_{\perp})}{\omega+i\tau_{\perp}^{-1}r_{\perp}}.
\label{transversek}
\end{equation}
This is a cubic equation in $\omega$.  

{\bf Short Times.}  An important solution in the time-domain is uniform decay of $v_{n}$ toward $v_{L}$.  Then for small $k$ the RHS of \eqref{transversek} is small if the second numerator is small, which leads to, with a correction for small finite $k$, 
\begin{equation}
\omega=-i\tau^{-1}_{\perp}(1+r_{\perp}) +i\frac{c^{2}k^{2}}{(1+r_{\perp})^{2}}\tau_{\perp}.
\label{uniformtransversedecay}
\end{equation}
Substitution of \eqref{transversek} into either \eqref{djdtA} or \eqref{dvLdtA} determines the normal mode structure; it gives the ratio $v_{n}/v_{F}$ as a function of $\omega$ or $k$. 

{\bf Low Frequency.} At low frequency ($\omega\tau_{\perp}\ll1$), eq.~\eqref{transversek} gives 
\begin{equation}
\omega\approx \pm c_{\perp}k(\frac{r_{\perp}}{1+r_{\perp}})^{1/2}-i\frac{c^{2}_{\perp}k^{2}\tau_{\perp}}{2(1+r_{\perp})^{2}}.
\label{lowomegaperp}
\end{equation}
Here $v_{n}$ and $v_{L}$ very nearly move together, with the leading term in \eqref{lowomegaperp} arising from \eqref{djdtB2} for $v_{n}=v_{L}$.  Likely diffusion terms give an additional term in $-ik^{2}$.  

{\bf High Frequency.} For high frequency ($\omega\tau_{\perp}\gg1$), eq.~\eqref{transversek} gives
\begin{equation}
k^{2}\approx \frac{\omega}{c^{2}_{\perp}}(\omega+ i\tau_{\perp}^{-1})=\frac{\omega^{2}}{c^{2}_{\perp}}(1+ i\frac{1}{\omega\tau_{\perp}}).
\label{transversek-high}
\end{equation}
This mode has small $v_{n}/v_{L}$, with roots given by
\begin{equation}
\omega=\pm c_{\perp}k-\frac{i}{2}\tau^{-1}_{\perp}.  
\label{highomegaperp}
\end{equation}
Neglecting dissipation this follows from \eqref{dvLdtB2} if $v_{n}$ is ignored.  Dissipation arises from friction of $v_{L}$ against $v_{n}$.  

\section{Longitudinal Modes ($\parallel$)}
\label{s:longitudinalmot}
We give all oscillating thermodynamic quantities primes: $\rho'$, $P'$, $\mu'$, $S'$, $T'$.   Only two of these variables are independent; we take them to be $\rho'$ and $S'$. 

In the linear regime, when a thermodynamic quantity multiplies a velocity, we may treat the thermodynamic quantity as a constant.  Because the motion is longitudinal we drop the subscript on the vector variables.  To make all the above equations of motion contain $\omega^{2}$, as necessary we take $\partial_{t}\rightarrow-i\omega$ and $\partial_{i}\rightarrow ik\delta_{iz}$ (thus $x$ and $y$ are transverse).  To relate $\omega$ and $k$ we will obtain two distinct relations proportional to $\rho'$ and $S'$.  

Then, with $K_{\parallel}=K+\tfrac{4}{3}\bar{\mu}$, 
\begin{eqnarray}
0&=&-\omega^{2}{\rho}' +\omega k j, 
\label{drhodt1L2}\\
0&=&-\omega^{2}j +K_{\parallel}k^{2}v_{L}. 
\label{djdt1L2}\\
0&=&-\omega^{2}v_{s} +\omega k\mu', 
\label{dvsdt1L2}\\
0&=&-\omega^{2}v_{L} +\omega k\mu'+ \frac{K_{\parallel}}{\rho_{L\parallel}}k^{2}v_{L}  -i\omega \tau_{\parallel}^{-1}(v_{L}-v_{n}), \qquad
\label{dvLdt1L2}\\
0&=&-\omega^{2} S' + \omega kSv_{n}. 
\label{dSdt1L2}
\end{eqnarray}

Since \eqref{drhodt1L2} gives $j=(\omega/k)\rho'$ and \eqref{djdt1L2} gives $v_{L}=(\omega^{2}/K_{\parallel}k^{2})j$, we have $v_{L}=(\omega^{3}/K_{\parallel}k^{3})\rho'$.  Further, \eqref{dvsdt1L2} gives $v_{s}=(k/\omega)\mu'$.  Thus we can eliminate all vectors in terms of thermodynamic scalars.  
Then, with the ratio $r_{\parallel}$ and longitudinal elasticity velocity $c_{\parallel}$ given by
\begin{equation}
r_{\parallel} \equiv \frac{\rho_{L\parallel}}{\rho_{n\parallel}}, \quad c^{2}_{\parallel}\equiv\frac{K_{\parallel}}{\rho_{L\parallel}},  
\label{cpar}
\end{equation}
using the above relations to replace various vectors by appropriate scalars, 
\eqref{dvLdt1L2} becomes
\begin{eqnarray}
0=(-\omega^{2}+c^{2}_{\parallel}k^{2}-i\frac{\omega}{\tau_{\parallel}})\frac{\omega^{3}\rho'}{ \rho_{L\parallel}c^{2}_{\parallel} k^{3}}
+\omega k\mu' 
+i\frac{\omega}{\tau_{\parallel}}(\frac{\omega}{k})\frac{S'}{S}. \quad
\label{dvLdt1L2Fa}
\end{eqnarray}

We now define the velocities $c_{\rho}$ and $c_{S}$ by
\begin{equation}
c^{2}_{\rho}\equiv\rho\frac{\partial\mu}{\partial \rho}>0, \quad c^{2}_{S}\equiv-S\frac{\partial\mu}{\partial S}\approx \frac{S^{2}}{\rho}\frac{\partial T}{\partial S}>0.
\label{crhocS}
\end{equation}
Then, writing $\mu'$ in terms of $\rho'$ and $S'$,   
\eqref{dvLdt1L2Fa}, on multiplication by $k/\omega$ and rearranging, yields a first relation: 
\begin{eqnarray}
&{\hskip-0.25cm}\Big[&{\hskip-0.25cm}(-\omega^{2}+c^{2}_{\parallel}k^{2}-i\frac{\omega}{\tau_{\parallel}})\frac{\omega^{2}}{\rho_{L\parallel}c^{2}_{\parallel}k^{2}}
+ k^{2} \frac{c^{2}_{\rho}}{\rho}\Big]\rho' 
{\hskip-0.1cm}={\hskip-0.1cm}-\Big[i\frac{\omega}{\tau_{\parallel}}- k^{2} c^{2}_{S} \Big] \frac{S'}{S}. \qquad
\label{dvLdt1L2Fb}
\end{eqnarray}

We now write $j$ from \eqref{j2} as
\begin{eqnarray}
j=\rho_{s_{\parallel}}(\frac{k}{\omega})\mu' 
+ \rho_{n_{\parallel}}(\frac{\omega}{k})\frac{S'}{S}
+ (\frac{\omega^{3}}{c^{2}_{\parallel} k^{3}})\rho'. \quad 
\label{j3}
\end{eqnarray}
As a result, \eqref{drhodt1L2} for $\partial_{t}j_{i}$ becomes 
\begin{eqnarray}
0=-\omega^{2}{\rho}' +\rho_{s_{\parallel}}k^{2}\mu' 
+ \rho_{n_{\parallel}}\omega^{2}\frac{S'}{S} 
+(\frac{\omega^{4}}{c^{2}_{\parallel} k^{2}})\rho'. \quad
\label{drhodt1L2a}
\end{eqnarray}

Writing $\mu'$ in terms of $\rho'$ and $S'$ and rearranging \eqref{drhodt1L2a} yields a second relation
\begin{eqnarray}
\Big[ \omega^{2}\rho_{n_{\parallel}}-\rho_{s_{\parallel}}c^{2}_{S}k^{2} \Big] \frac{S'}{S} 
=\Big[\omega^{2} -\frac{\rho_{s_{\parallel}}}{\rho}c^{2}_{\rho}k^{2} - \frac{1}{c^{2}_{\parallel}}\frac{\omega^{4}}{k^{2}} \Big]\rho'. \quad
\label{drhodt1L2c}
\end{eqnarray}

Cross-multiplying \eqref{drhodt1L2c} and \eqref{dvLdt1L2Fb} gives the implicit dispersion relation 
\begin{eqnarray}
\Big[ \rho_{n_{\parallel}}\omega^{2} -\rho_{s_{\parallel}}c^{2}_{S}k^{2} \Big]
\Big[(-\omega^{2}+c^{2}_{\parallel}k^{2}-i\frac{\omega}{\tau_{\parallel}})(\frac{\omega^{2}}{ \rho_{L}c^{2}_{\parallel} k^{2}})+ \frac{k^{2}}{\rho} c^{2}_{\rho} \Big]\cr
  =-\Big[i\frac{\omega}{\tau_{\parallel}}- k^{2} c^{2}_{S} \Big] \Big[\omega^{2} -\frac{\rho_{s_{\parallel}}}{\rho} c^{2}_{\rho}k^{2} - \frac{1}{c^{2}_{\parallel}}\frac{\omega^{4}}{k^{2}}
 \Big] .\qquad\quad
\label{aux20}
\end{eqnarray}

{\bf Short Times.}  
The drag term between uniform $v_{n}$ and $v_{L}$ ($k=0$) gives the short-time response 
where $v_{n}$ and $v_{L}$ come into motion together.  For $k=0$ \eqref{aux20} yields 
\begin{eqnarray}
\omega=-i\tau_{\parallel}^{-1}(1+\frac{\rho_{L\parallel}}{\rho_{n\parallel}})=-i\tau_{\parallel}^{-1}(1+r_{L\parallel}). \quad
\label{aux20b}
\end{eqnarray}
This represents uniform decay of $v_{n}$ relative to $v_{L}$.  It has the same form as \eqref{uniformtransversedecay} for uniform transverse decay. 

In what follows we drop the tensor index $\parallel$ on $\tau$, $\rho_{s}$, $\rho_{n}$, and $\rho_{L}$, but not on $c_{\parallel}$.

{\bf Low Frequency.} At low frequency $\omega\tau\ll 1$, 
we expect that drag between $v_{n}$ and $v_{L}$ dominates, so we expect $v_{n}$ and $v_{L}$ to move together as one degree of freedom.  With $v_{s}$ as another degree of freedom, we then expect two coupled modes.  In \eqref{aux20} we treat the terms in $\tau^{-1}$ as much larger than $\omega$, $c_{S}k$, $c_{\parallel}k$, and $c_{\rho}k$.  Factoring out the leading terms in $-i\omega/\tau$ then leads to 
\begin{eqnarray}
\Big[ \rho_{n}\omega^{2} -\rho_{s}c^{2}_{S}k^{2} \Big] \Big[\frac{\omega^{2}}{ \rho_{L}c^{2}_{\parallel} k^{2}} \Big] 
= \Big[\omega^{2} -\frac{\rho_{s}}{\rho} c^{2}_{\rho} k^{2} - \frac{1}{c^{2}_{\parallel}}\frac{\omega^{4}}{k^{2}} \Big]. \quad
\label{aux20d}
\end{eqnarray}
With the effective mass associated with the lattice $\rho_{Ln}\equiv \rho_{L}+\rho_{n}$, this can be rewritten as 
\begin{eqnarray}
0&=&\omega^{4}-\omega^{2}k^{2}\Big[\frac{\rho_{s}}{\rho_{Ln}}c^{2}_{S} + \frac{\rho_{L}}{\rho_{Ln}}c^{2}_{\parallel}\Big]
+k^{4} \frac{\rho_{s}}{\rho}\frac{\rho_{L}}{\rho_{Ln}}c^{2}_{\rho}c^{2}_{\parallel}. \quad
\label{lowomeigeq2}
\end{eqnarray}

{\bf High Frequency.}   A high frequency ($\omega\tau\gg 1$), 
drag between $v_{n}$ and $v_{L}$ is negligible, so each velocity variable is independent. 
The high frequency limit of \eqref{aux20} eliminates the ten imaginary terms and gives 
\begin{eqnarray}
&&\Big[ \rho_{n}\omega^{2} -\rho_{s}c^{2}_{S}k^{2} \Big] 
\Big[(-\omega^{2}+c^{2}_{\parallel}k^{2})(\frac{\omega^{2}}{ \rho_{L}c^{2}_{\parallel} k^{2}})+ \frac{k^{2}}{\rho} c^{2}_{\rho} \Big]\cr
&&  =\Big[\omega^{2} -\frac{\rho_{s}}{\rho} c^{2}_{\rho} k^{2} - \frac{1}{c^{2}_{\parallel}}\frac{\omega^{4}}{k^{2}}
 \Big] \Big[k^{2} c^{2}_{S} \Big].\qquad\quad
\label{aux21}
\end{eqnarray}
This has a term cubic in $\omega^{2}$, but the terms in $\omega^{0}$ cancel, so it has a $\omega=0$ solution and two others.  

For $\omega=0$, both \eqref{dvLdt1L2Fb} and \eqref{drhodt1L2c} lead to $\mu'=0$, so even with drag $\omega=0$ is an eigenfrequency with eigenfunction $\mu'=0$.  For this mode $j=0$ and $v_{n}=0$, but $v_{s}$ and $v_{L}$ are non-zero and do not accelerate, subject to $\rho_{s}v_{s}+\rho_{L}v_{L}=0$.  Likely when when nonlinear drag terms are included this mode will become dissipative.  We do not consider it further.  

With $\rho_{Ls}\equiv \rho_{L}+\rho_{s}$, \eqref{aux21} can be rewritten as
\begin{eqnarray}
0&=&\omega^{4}-\omega^{2}k^{2}\Big[\frac{\rho_{Ls}}{\rho_{n}} c^{2}_{S} +c^{2}_{\parallel} \Big]
+k^{4}c^{2}_{\parallel}\Big[\frac{\rho_{Ls}}{\rho_{n}} c^{2}_{S}
-  \frac{\rho_{L}}{\rho} c^{2}_{\rho}\Big]. \qquad\quad 
\label{aux23}
\end{eqnarray}
For $\rho_{L}=0$ the two roots involve the elastic velocity $c_{\parallel}$ and the second sound velocity $c_{2}=\sqrt{(\rho_{s}/\rho_{n})c_{S}^{2}}$.  For a superfluid 
two roots are the first sound velocity and the second sound velocity.  The difference is that the solid has elastic restoring forces rather than pressure restoring forces.  For $\rho_{L}\ne 0$ the roots must be obtained by solving \eqref{aux21}, with no simple interpretation other than what can be obtained by studying the eigenmodes, as in \eqref{drhodt1L2c}. 

For both roots to be positive, the second bracket in \eqref{aux23} must be positive. 
If $c_{1}$ is the first sound velocity, then for a phonon-dominated system at low $T$ the first term in the second bracket 
is $\tfrac{1}{3}c^{2}_{1}$ and $c_{\rho}\approx c_{1}$.  Also, at low $T$ we have $\rho_{Ls}\approx\rho$, so the largest value of $\rho_{L}\rho_{s}\approx \rho_{L}(\rho-\rho_{L})$ is $\tfrac{1}{4}\rho^{2}$.  Therefore, at least at low $T$, the bracket is in the range $\tfrac{1}{3}$ to $\tfrac{1}{12}$ multiplied by $c^{2}_{1}$, so the system is stable.  


For both the longitudinal and the transverse modes, as the frequency increases there will be a cross-over from low to high frequency behavior.  This might appear in acoustics or acoustics-related experiments.\cite{Lev21}
\section{Summary and Conclusions}
By Landau\cite{Landau41} and by Leggett,\cite{Leggett70} at $T=0$ the macroscopic two-velocity theory of AL for a supersolid has missing inertial mass. We have added a new momentum-bearing velocity $v_{L}$ associated with the mass density, $\rho_{L}\equiv\rho-\rho_{s}-\rho_{n}$, which gives an additional term to the thermodynamics and requires a new dynamical equation. 

Using Onsager's irreversible thermodynamics we have derived the dynamical equations.  Whereas in AL $\dot{u}$ and $v_{n}$ equilibrate by diffusion, in the present work $\dot{u}$ and $v_{L}$ (rather than $v_{n}$) equilibrate by diffusion.  Further, $v_{L}$ and $v_{n}$ tend to equilibrate frictional drag, although $v_{L}$ is explicitly subject to additional forces. 

The theory indicates that the supersolid component (the lattice) is driven by not only the elastic force, but also frictional drag relative to the normal fluid, and by the negative gradient of the chemical potential per mass.  This latter force is precisely as for the superfluid component and indicates that the supersolid component may indeed be considered part of the ground state.  

At high frequencies relative to $\tau^{-1}$, there are three longitudinal modes and two doubly degenerate transverse modes.  At low frequencies relative to $\tau^{-1}$, because drag tends to make $v_{L}$ and $v_{n}$ act as a single degree of freedom, there are two longitudinal modes and one 
transverse mode.  As a function of frequency there is a crossover in the response, 
where for $\omega\tau\ll1$ drag causes $v_{n}$ and $v_{L}$ to move together, and for $\omega\tau\gg1$ they are independent. 

Note that the atomically-varying {\it microscopic,} superfluid velocity is defined relative to the lattice velocity $v_{L}$, which can engage in shear motion.\cite{Leggett70}  However, in AL and in the present work the {\it macroscopic} superfluid velocity is taken to be shear-free.  (A macroscopic average over the microscopic non-Galilean $v_{si}$ gives the macroscopic non-Galilean $\bar{v}_{si}$.  By adding a phase gradient corresponding to a boost $v_{si}$ becomes a Galilean velocity.  Our earlier work on both the microscopic and macroscopic theory,\cite{Saslow77} using the two-velocity framework of AL, first appropriately used a non-Galilean microscopic $v_{si}$ to study a microscopic response function, and then appropriately used a macroscopic Galilean $v_{si}$ to derive nonlinear equations that reduced to those of AL.  However, in applying the theory to obtain the normal modes, Ref.~\onlinecite{Saslow77} inappropriately reverted to a non-Galilean $v_{si}$ defined relative to $v_{Li}$.  We find  that such a non-Galilean $v_{si}$ leads to dissipation under uniform superflow, and for that reason is considered inappropriate.\cite{WMSUnpublished} \\ 

Gross-Pitaevskii\cite{Gross61,Pitaevskii61} theory (GP) can be used to give the excitations in superfluid atomic gases.\cite{PitaString,PethickSmithGases}  Irreversible thermodynamics cannot do this.  Also, GP theory gives modes associated with density (Higgs-like) and phase (superfluid-like); and with lattice ordering can give solid-like modes.\cite{YooDorsey10}  If the frequency of the Higgs mode is below the presumable rapid equilibration rate needed for irreversible thermodynamics to hold, then by adding an amplitude degree of freedom to the irreversible thermodynamics such a mode might be describable by irreversible thermodynamics.  However, to our knowledge GP theory (and other pure field theories) cannot include the effect of near-equilibrium thermal excitations that drift with $v_{ni}$ to provide the momentum of the normal fluid.\cite{Turski}  Thus GP theory cannot yield second sound in an ordinary superfluid.  The utility of irreversible thermodynamics is that, by using the rate of entropy production, it can treat the effect of the thermal excitations.  

Thus both irreversible thermodynamics and Gross-Pitaevskii theory have strengths and limitations. 

I would like to acknowledge valuable correspondence and conversations with Moses Chan and Mario Liu. 



{}

\appendix
\section{Irreversible Thermodynamics}
\label{app:IrTh}

\subsection{Rate of Heat Production}
\label{ss:Heat}
To find the density of the heating rate $R$ we employ \eqref{dSdt1}, an integration by parts, the energy differential \eqref{dE123}, energy conservation \eqref{Edot}, and the equations of motion in Sect.~\ref{s:Dynamics}.  Then 
\begin{eqnarray}
R&=&T\dot{S}+T\partial_{i}\Big[Sv_{ni}+q_{i}/T\Big]\cr
&=&T\dot{S}+\partial_{i}\Big[TSv_{ni}+q_{i}\Big]-(Sv_{ni}+q_{i}/T)\partial_{i}T \cr
&=& \dot{E}-\lambda_{ik}\partial_{i}U_{k}-\mu \dot{\rho}-v_{ni} \partial_{t}j_{i}-j_{si}\dot{v}_{si}\cr
&&-j_{Li}\dot{v}_{Li}+\partial_{i}\Big[TSv_{ni}+q_{i}\Big]-(Sv_{ni}+q_{i}/T)\partial_{i}T \cr
&=&-\partial_{i}Q_{i}-\lambda_{ik}\partial_{i}U_{k}+\mu\partial_{i}j_{i}+v_{ni}\partial_{k}\Pi_{ik}+j_{si}\partial_{i}\varphi \cr
&&- j_{Li}Y_{i}+\partial_{i}\Big[TSv_{ni}+q_{i}\Big]-(Sv_{ni}+q_{i}/T)\partial_{i}T. \qquad
\label{R1}
\end{eqnarray}

On the right-hand-side (RHS) we collect the divergence terms and integrate by parts on the terms $-\lambda_{ik}\partial_{i}U_{k}$, $v_{ni}\partial_{k}\Pi_{ik}$, and $j_{si}\partial_{i}\varphi$.  Then 
\begin{eqnarray}
R&=&-\partial_{i}\Big[Q_{i}-TSv_{ni}-q_{i}+\lambda_{ik}U_{k}-v_{nk}\Pi_{ki}-j_{si}\varphi \Big]\cr
&&+U_{k}\partial_{i}\lambda_{ik}+\mu\partial_{i}(\rho v_{ni}+j_{si}+j_{Li})-\Pi_{ik}\partial_{k}v_{ni} \cr 
&&-\varphi\partial_{i}j_{si} - j_{Li}Y_{i}-(Sv_{ni}+q_{i}/T)\partial_{i}T. \quad\qquad
\label{R2a}
\end{eqnarray}
The terms $\mu\partial_{i}(j_{si}+j_{Li})$ cause both $j_{si}$ and $j_{Li}$ to couple to $\mu$.  The term in $j_{Li}$ did not appear in previous theories of the supersolid. 

Integration by parts on the terms $\mu \partial_{i}(\rho v_{ni}+j_{Li})$ gives 
\begin{eqnarray}
R&=&-\partial_{i}\Big[Q_{i}-TSv_{ni}-q_{i}+\lambda_{ik}U_{k}-v_{nk}\Pi_{ki}\cr
&& \qquad  -j_{si}\varphi - \mu \rho v_{ni} -j_{Li}\mu \Big] \cr
&&+U_{k}\partial_{i}\lambda_{ik}-\rho v_{ni}\partial_{i}\mu -\Pi_{ik}\partial_{k}v_{ni} \cr
&& +(\mu-\varphi)\partial_{i}j_{si}  -(\partial_{i}\mu+Y_{i})j_{Li} \cr
&&-(Sv_{ni}+q_{i}/T)\partial_{i}T. 
\label{R2}
\end{eqnarray}

For $v_{Li}$ to respond to elasticity we rearrange to have $j_{Li}$ multiplied by an elasticity term.  Thus, to $R$ we add the identity 
$$0=-v_{nk}\partial_{i}\lambda_{ik}+\partial_{i}(v_{nk}\lambda_{ik})-\lambda_{ik}\partial_{i}v_{nk};$$
we later employ $v_{Lk}-v_{nk}=\rho_{L}^{-1}j_{Lk}$.  The first term of the identity goes with $U_{k}\partial_{i}\lambda_{ik}$, the second with the divergence, and the third with $-\Pi_{ik}\partial_{k}v_{ni}$.  We also combine $-\rho\mu v_{ni}$ and $-TSv_{ni}$ within the divergence, and combine $v_{ni}\rho\partial_{i}\mu$ and $v_{ni}S\partial_{i}T$ outside the divergence.  Then 
\begin{eqnarray}
R&=&-\partial_{i}\Big[Q_{i}-q_{i}+\lambda_{ik}U_{k}-v_{nk}(\Pi_{ki}+\lambda_{ik}) \cr
&& {\hskip-.15cm} -v_{ni}(TS+\rho\mu) -j_{si}\varphi -j_{Li}\mu  \Big] \cr
&&+[(U_{k}-v_{Lk})+(v_{Lk}-v_{nk})]\partial_{i}\lambda_{ik} -(\Pi_{ik}+\lambda_{ki})\partial_{k}v_{ni} \cr
&& +(\mu-\varphi)\partial_{i}j_{si} -(\partial_{i}\mu+Y_{i})j_{Li} \cr
&&-v_{ni}(\rho \partial\mu+S\partial_{i}T)-(q_{i}/T)\partial_{i}T.
\label{R3a}
\end{eqnarray}

We now employ $v_{Lk}-v_{nk}=\rho_{L}^{-1}j_{Lk}$.  (The tensor version is $v_{Lk}-v_{nk}=\rho_{L,ki}^{-1}j_{Li}$.)  Then 
\begin{eqnarray}
R
&=& -\partial_{i}\Big[Q_{i}-q_{i}+\lambda_{ik}U_{k}-v_{nk}(\Pi_{ki}+\lambda_{ki})\cr
&& \qquad -(TS+\rho\mu)v_{ni}  -j_{si}\varphi -j_{Li}\mu \Big] \cr
&& +(U_{k}-v_{Lk})\partial_{i}\lambda_{ik} -(\Pi_{ik}+\lambda_{ki})\partial_{k}v_{ni} \cr
&& +(\mu-\varphi)\partial_{i}j_{si}  +[\rho_{L}^{-1}\partial_{k}\lambda_{ki}-(\partial_{i}\mu+Y_{i})]j_{Li} \cr 
&& -v_{ni}(\rho \partial\mu+S\partial_{i}T) -(q_{i}/T)\partial_{i}T.
\label{R4lin}
\end{eqnarray}  
This contains thermodynamic ``forces'' involving gradients of $T$, $\lambda_{ki}$, $v_{ni}$ and $v_{si}$, and the current $j_{Li}$, as indicated earlier.  The term in $-v_{ni}\rho\partial_{i}\mu$ has been combined with the term $-v_{ni}S\partial_{i}T$.  

Using the linearized version of the Gibbs-Duhem relation we will replace the last two terms by the single term $-v_{ni}\partial_{i}P$.  On integrating by parts, that term will be absorbed by the divergence and by the term $-(\Pi_{ik}+\lambda_{ki})\partial_{k}v_{ni}$. 

\subsection{Linearized Rate of Heat Production $R$}
\label{ss:HeatLinearized}
In the last line of \eqref{R4lin} two terms become, on using the linearized Gibbs-Duhem relation \eqref{G-Duh123}, given by $dP=SdT+\rho d\mu+\dots$  (with dots for the higher order terms),  
\begin{eqnarray}
&&{\hskip-1.4cm}-v_{ni}(S\partial_{i}T+\rho \partial_{i}\mu)=-v_{ni}\partial_{i}P+\dots \cr
&=&[-\partial_{i}(Pv_{ni})+P\partial_{i}v_{ni}]
+[v_{ni}w_{jk}\partial_{i}\lambda_{jk}\cr
&&+v_{ni}j_{k}\partial_{i}v_{nk}+v_{ni}v_{sk}\partial_{i}j_{sk}+v_{ni}v_{Lk}\partial_{i}j_{Lk}].
\label{aux1}
\end{eqnarray}
After the second equality, the two bracketed terms involving $P$ are of the desired form.  These are all that are needed to obtain the lowest order terms in the dynamical equations, which we proceed to obtain.  

Using \eqref{aux1}, eq.~\eqref{R4lin} becomes
\begin{eqnarray}
R  
&=& -\partial_{i}\Big[Q_{i}-q_{i}+\lambda_{ik}U_{k}-v_{nk}(\Pi_{ki}+\lambda_{ki})\cr
&& \qquad -(TS+\rho\mu-P)v_{ni} - j_{si}\varphi - j_{Li}\mu \Big] \cr
&& +(U_{k}-v_{Lk})\partial_{i}\lambda_{ik} -(\Pi_{ik}+\lambda_{ki}-\delta_{ik}P)\partial_{k}v_{ni} \cr 
&& +(\mu-\varphi)\partial_{i}j_{si}  +[\rho_{L}^{-1}\partial_{k}\lambda_{ki}-(\partial_{i}\mu+Y_{i})]j_{Li} \cr 
&& -(q_{i}/T)\partial_{i}T +\dots \qquad
\label{R4lina}
\end{eqnarray} 
Thus the thermodynamic forces that appear are $\partial_{i}\lambda_{ik}$, $\partial_{k}v_{ni}$, $\partial_{i}j_{si}$, $j_{Li}$, and $\partial_{i}T$.  Other than the term in $j_{Li}$, the other terms appear either in theories for ordinary solids or for an ordinary fluid or superfluid.  

The only new term involves $j_{Li}$, where the source term $Y_{i}$ with units of force per mass density, must be determined. 
$Y_{i}$ is a real-space vector with units of force density per mass.  In addition to being subject to lattice elasticity and the back reaction of the superfluid, it can have dissipative terms proportional to the thermodynamic forces.  Of these, 
$\partial_{i}\lambda_{ik}$ and $\partial_{i}T$ have time-reversal symmetries that are opposite that of $j_{Lk}$, giving contributions to $R$ that are odd under time-reversal.  Thus they are {\it not}  dissipative.  $\partial_{k}v_{ni}$, $j_{Li}$, and $\partial_{ik}j_{si}$ have time-reversal symmetries that are the same as that of $j_{Lk}$, and thus {\it are} dissipative.  
The next section shows that the term in $j_{Lk}$ gives drag of the lattice $v_{Lk}$ against the normal fluid $v_{nk}$. 

Using \eqref{R4lina} we now write down the products of the fluxes multiplying the thermodynamic forces, neglecting nonlinear terms.  

\subsection{Linearized Thermodynamic Fluxes}
\label{ss:FluxesLinearized}
The flux $Q_{i}$ may be obtained by setting the divergence term in \eqref{R4lin} to zero.  This gives $Q_{i}$ in terms of other quantities, some of which are fluxes to be determined.  
\begin{eqnarray}
Q_{i}&=&q_{i}-\lambda_{ik}U_{k}+v_{nk}(\Pi_{ki}+\lambda_{ki})+j_{si}\varphi  +j_{Li}\mu \cr
&&+(TS+\rho\mu-P)v_{ni}.
\label{Qi}
\end{eqnarray}
We will not consider $Q_{i}$ further.  

{\bf (a)} For $\partial_{i}\lambda_{ik}$, the dissipation term in \eqref{R4lina} is 
\begin{eqnarray}
(U_{k}-v_{Lk})\partial_{i}\lambda_{ik}\equiv -F_{k}\partial_{i}\lambda_{ik},  
\label{dlambdaterms}
\end{eqnarray}
where $F_{i}$ is the ``flux'' associated with $\partial_{i}\lambda_{ik}$.  The diagonal dissipative term in $F_{k}$ is $\sim \partial_{i}\lambda_{ik}$.  

More completely, introducing diffusion coefficients using the notation of AL, the irreversible thermodynamics gives flux terms proportional to the ``forces'', subject to space, time, and rotational symmetry requirements.  Then by symmetry 
\begin{equation}
F_{i} \equiv v_{Li}-U_{i}=-\beta_{ik}\partial_{l}\lambda_{kl}-\alpha_{ik}\partial_{k}T.
\label{Fk2a}
\end{equation}
Thus the rate of dissipation $R$ of \eqref{R4lin} contains the terms
$$(\beta_{ik}\partial_{l}\lambda_{kl} +\alpha_{ik}\partial_{k} T)\partial_{i}\lambda_{ik}.$$

Without dissipation $F_{i}=0$, so 
\begin{equation}
U_{k}=v_{Ek}=v_{Lk}; \hbox{ (no dissipation)}
\label{vLkReactive}
\end{equation}
thus the lattice displacement vector and the lattice mass move together.  Since this gives the dynamics of $u_{i}$, which can only be diffusive, we conclude that without dissipation there is no diffusion. 

{\bf (b)} For $\partial_{i}v_{nk}$, the dissipation term in \eqref{R4lina} is 
\begin{equation}
-(\Pi_{ik}+\lambda_{ki}-P\delta_{ik})\partial_{k}v_{ni}\equiv -\pi_{ik}\partial_{k}v_{ni},
\label{dvnterms0}
\end{equation}
where by symmetry, in the notation of AL 
\begin{equation}
\pi_{ik} \equiv \Pi_{ik}+\lambda_{ki}-P\delta_{ik}=-\eta_{iklm}\partial_{m} v_{nl}-\zeta_{ik}\partial_{l}j_{sl}. 
\label{piik2a}
\end{equation}
Thus the rate of dissipation $R$ of \eqref{R4lin} contains the terms
$$(\eta_{iklm}\partial_{m} v_{nl}+\zeta_{ik}\partial_{l}j_{sl})\partial_{k}v_{ni}.$$

Without dissipation $\pi_{ik}=0$, so we have 
\begin{equation}
\Pi_{ik}=-\lambda_{ki}+P\delta_{ik}.  \hbox{ (no dissipation)}
\label{PikReactive}
\end{equation}

{\bf (c)} For $j_{si}$ the dissipation term in \eqref{R4lina} is 
\begin{eqnarray}
-(\mu-\varphi)\partial_{i}j_{si}\equiv-\psi \partial_{i}j_{si}. 
\label{djsterms}
\end{eqnarray}
By AL (12), we have
\begin{equation}
\psi \equiv \varphi-\mu=-\zeta_{ik}\partial_{k}v_{ni}-\chi\partial_{i}j_{si}.
\label{psiReactive}
\end{equation}
Thus the rate of dissipation $R$ of \eqref{R4lin} contains the term
$$(\zeta_{ik}\partial_{k}v_{ni}+\chi\partial_{i}j_{si})\partial_{k}j_{sk}.$$
By the Onsager principle that dissipative heating terms in $R$ are the same if flux and force are interchanged, we have
$$\zeta_{ik}=\zeta_{ki}.$$

Without dissipation $\psi=0$, and for local equilibrium $\phi=\mu$, so 
\begin{equation}
\dot{v}_{si}=-\partial_{i}\varphi=-\partial_{i}\mu.  \hbox{ (no dissipation)}
\label{vsdotReactive}
\end{equation}

{\bf (d)} For $j_{Li}$, a flux that does not appear in AL, the dissipation term in \eqref{R4lina}  is
\begin{eqnarray}
[\rho_{L}^{-1}\partial_{k}\lambda_{ki} -(\partial_{i}\mu+Y_{i}) ] j_{Li} \equiv -\psi_{i}j_{Li}.  
\label{dvLterms}
\end{eqnarray}
where $\psi_{i}$ also does not appear in AL.  

Then, including dissipation from drag -- and neglecting diffusion terms, which are second order in gradients -- we introduce the drag tensor $C_{ik}$ by 
\begin{eqnarray}
\psi_{i}& \equiv & Y_{i}+\partial_{i}\mu-\rho_{L}^{-1}\partial_{k}\lambda_{ki} =-C_{ik}j_{Lk} \cr
&=&-C_{ij}\rho_{Ljk}(v_{Lk}-v_{nk})\equiv -\tau^{-1}_{ik}(v_{Lk}-v_{nk}). \qquad
\label{Ui2a}
\end{eqnarray}
Eq.~\eqref{Ui2a} defines a tensor relaxation rate $\tau^{-1}_{ik}$ that, when we consider an isotropic solid, will have only longitudinal and (degenerate) transverse components.  Onsager symmetry gives $C_{ik}=C_{ki}$. 

Without dissipation $\psi_{i}=0$, so 
\begin{equation}
\dot{v}_{Li}=Y_{i}=-\partial_{i}\mu+\rho_{L}^{-1}\partial_{k}\lambda_{ki}. \hbox{ (no dissipation)}
\label{Ui2b}
\end{equation}
Thermodynamics gives $\mu$, and elasticity gives $\lambda_{ki}$.  
Note that both $\dot{v}_{Li}$ and $\dot{v}_{si}$ have acceleration terms $-\partial_{i}\mu$.

We have neglected diffusion terms in $\partial_{i}v_{Lk}$, $\partial_{i}v_{nk}$, and $\partial_{i}v_{si}$.  By Onsager symmetry, the equations for $j_{i}$ and $v_{si}$ then would contain additional terms in $\partial_{i}v_{Lk}$.  

{\bf (e)} For $\partial_{i} T$, there is only the dissipation term in \eqref{R4lina}, given by the usual  
\begin{equation}
-\frac{q_{i}}{T} \partial_{i} T. 
\label{dTterms}
\end{equation}
Then 
\begin{equation}
q_{i}=-\kappa_{ik} \partial_{k} T-\alpha_{ik} \partial_{l}\lambda_{kl}. 
\label{qia}
\end{equation}
By the Onsager principle for dissipative fluxes 
$$\alpha_{ik}=\alpha_{ki}.$$


\begin{thebibliography}{}

\bibitem{Landau41} L. D. Landau, J. Phys. USSR 5, 71 (1941).  Reprinted in I. M. Khalatnikov, ``Theory of Superfluidity'', Benjamin, New York (1965). 


\bibitem{AL69} A. F. Andreev and I. M. Lifshitz, Sov. Phys. JETP 29, 1107 (1969). ``Quantum Theory of Defects in Crystals.''  

\bibitem{OrdSolid} Even for ordinary systems, AL theory with vacancies and interstitials has yet to be systematically investigated.  Following AL, systems with different vacancy and interstitial content, but subject to the same applied pressure, will have the same net internal stress, but may have different internal pressures and elastic stresses.  M. Sears and W. M. Saslow, Phys. Rev. B 82, 134304 (2010).  ``Andreev-Lifshitz hydrodynamics applied to an ordinary solid under pressure.''  

\bibitem{Chester70} G.V. Chester, Phys. Rev. A 2, 256 (1970). ``Speculations on Bose-Einstein Condensation and Quantum Crystals''.

\bibitem{Leggett70} A. J. Leggett, Phys. Rev. Lett. 25, 1543 (1970).  ``Can a Solid Be `Superfluid'?''

\bibitem{macrocurrents} Such ground state atomic currents are stable because of the finite energy gap to excited states.  To be an insulator there  must be no delocalized states, by W. Kohn, Phys. Rev. 133, A171 (1964), ``Theory of the Insulating State''. 
If macroscopic currents do occur, then they ares stable only if something like the Landau criterion is satisfied.\cite{Landau41}

\bibitem{AmperianCurrent} Such dissipation-free atomic currents occur spontaneously in magnetic atoms and in the diamagnetic response of atoms.  They were proposed by Amp\`ere in 1821, when atoms were not universally accepted.  However, two unpublished papers on this subject, by Fresnel,  were much later found among Amp\`ere's manuscripts; the two were published in 1885.  See Amp\`ere's Electrodynamics, by A. K. T. Assis and J. P. M. C. Chaib, Apeiron, Montreal, CA (2015).  This work may be downloaded at https://www.ifi.unicamp.br/$\sim$assis/Amperes-Electrodynamics.pdf. 

\bibitem{FernandezPuma74} J. F. Fernandez and M. Puma, J. Low Temp. Phys. 17, 131-141. ``Superfluidity of Solid $^{4}$He.''


\bibitem{GRS07} D. E. Galli, L. Reatto, and W. M. Saslow, Phys. Rev. B 76, 052503 (2007), ``Bounds for the superfluid fraction from exact quantum Monte Carlo local densities.''

\bibitem{Saslow12} W. M. Saslow, J. Low Temp. Phys. 160, 248-263. ``On the Superfluid Fraction and the Hydrodynamics of Supersolids.''  See Sect. 4.7. 

\bibitem{Mullin71} W.J. Mullin, Phys. Rev. Lett. 26, 611 (1971). ``Cell Model of a Bose-Condensed Solid''.

\bibitem{KimChan04} E. Kim and M. H. W. Chan, Nature 427, 225-227 (2004).  ``Probable observation of a supersolid helium phase.''

\bibitem{KimChan12} D. A. Kim and M. H. W. Chan, Phys. Rev. Lett.109, 155301 (2012). ``Absence of supersolidity in solid helium in porous Vycor glass''. 

\bibitem{ChanReview14} M. H. W. Chan, R. B. Hallock, L. Reatto, J. Low Temp. Phys. 172, 317-363 (2013). ``Overview on Solid 4He and the Issue of Supersolidity''. 

\bibitem{Leonard17} J. L\'eonard {\it et al}, 
Nature 543, 87 (2017).  ``Supersolid formation in a quantum gas breaking a continuous translational symmetry.''

\bibitem{Li17} J.-R. Li {\it et al}, 
Nature 543, 91 (2017).  ``A stripe phase with supersolid properties in spin-orbit-coupled Bose-Einstein condensates.''

\bibitem{Tanzi19A} L. Tanzi {\it et al}, 
Phys. Rev. Lett. 122, 130405 (2019).  ``Observation of a dipolar quantum gas with metastable supersolid properties.'' 

\bibitem{Bottcher19} F. B\"ottcher {\it et al}, 
Phys. Rev. X 9, 011051 (2019).  ``Transient supersolid properties in an array of dipolar quantum droplets.''

\bibitem{Chomaz19} L. Chomaz {\it et al}, Phys. Rev. X 9, 021012 (2019).  ``Long-lived and transient supersolid behaviors in dipolar quantum gases.''

\bibitem{Tanzi19B} L. Tanzi {\it et al}, Nature 574, 382 (2019).  ``Supersolid symmetry breaking from compressional oscillations in a dipolar quantum gas.''

\bibitem{Guo19} M. Guo {\it et al}, Nature 574, 386 (2019).  ``The low-energy Goldstone mode in a trapped dipolar supersolid.''

\bibitem{Lev21} Y. Guo {\it et al}, Nature 599, 211 (2021).  ``An optical lattice with sound.''

\bibitem{Ferlaino24} E. Casotti {\it et al}, 
Nature 635, 327-331 (2024). ``Observation of vortices in a dipolar supersolid.''  This work contains a more complete list of citations than presented here. 


\bibitem{YooDorsey10} C.-D. Yoo and A. T. Dorsey, Phys. Rev. B 81, 134518 (2010), ``Hydrodynamic theory of supersolids: Variational principle, effective Lagrangian, and density-density correlation function''. 

\bibitem{Poli24} E. Poli, D. Baillie, F. Ferlaino, and P. B. Blakie, Phys. Rev. A 110, 053301 (2024), ``Excitations of a two-dimensional supersolid''. 

\bibitem{Onsager1} L. Onsager, Phys. Rev. 37, 405-426 (1931).  ``Reciprocal Relations in Irreversible Processes. I''.  This work earned Onsager the Nobel Prize in chemistry for rigorously relating, by non-equilibrium correlations, heat flow from voltage gradients to charge flow from temperature gradients.  The macroscopic theory for correlations of non-equilibrium heat and charge fluxes is given in Sect. 5, ``The Principle of the Least Dissipation of Energy''.   L. Onsager, Phys. Rev. 38, 2265-2279 (1931), ``Reciprocal Relations in Irreversible Processes. II''  discusses correlations of non-equilibrium processes due to arbitrary types of flux. 

\bibitem{FN2} The liquid to solid transition is usually discontinuous, or first-order. 




\bibitem{LLElasticity} L. D. Landau and E. M. Lifshitz, {\it Elasticity}, 2nd ed. (Pergamon, 1970).  See Sect. 22. 

\bibitem{inertialdrag} In Sect.~11 of L. D. Landau and E. M. Lifshitz, {\it Fluid Mechanics}, 2nd ed. (Pergamon, 1987), the term ``drag'' means lossless inertial drag, not dissipative frictional drag. 







\bibitem{FN4} In AL, the equation for $\dot{u}_{i}$ is not written down explicitly.  The term in $v_{Li}$ is new.  






















\bibitem{WMSUnpublished} W. M. Saslow, unpublished. 

\bibitem{Saslow77} W. M. Saslow, Phys. Rev. B 15, 173 (1977). ``Microscopic and hydrodynamic theory of superfluidity in periodic solids.''

\bibitem{Gross61} E.P. Gross, Nuovo Cimento 20, 454-457 1961).  ``Structure of a quantized vortex in boson systems.''

\bibitem{Pitaevskii61} L.P. Pitaevskii, Sov. Phys. JETP 13,  451-454 (1961).  ``Vortex lines in an imperfect Bose gas.''

\bibitem{PitaString} L. Pitaevskii and S. Stringari, Bose-Einstein Condensation and Superfluidity, 2nd Edition (Oxford University Press, 2016).

\bibitem{PethickSmithGases} C. J. Pethick and H. Smith, Bose-Einstein Condensation in Dilute Gases, 2nd Edition (Cambridge University Press, 2016).   

\bibitem{Turski} L. A. Turski and K. Pawlowski, Phys. Lett. A 381, 1710-1713 (2017).  "On the dissipative version of the Gross-Pitaevski equation."














\end{thebibliography}
\end{document}